\documentclass[letterpaper,conference]{IEEEtran}
\pdfoutput=1

\usepackage[hyphens]{url}

\usepackage{cite}

\ifCLASSINFOpdf
   \usepackage[pdftex]{graphicx}
 \DeclareGraphicsExtensions{.pdf,.jpeg,.png}
\else
  \DeclareGraphicsExtensions{.eps}
\fi

\usepackage{listings}
\usepackage{booktabs} 
\usepackage{algorithm2e}
\usepackage{multirow}

\usepackage{url}

\begin{document}
%
\title{OSDF: An Intent-based Software Defined Network Programming Framework}

\author{\IEEEauthorblockN{Douglas Comer\IEEEauthorrefmark{1}, Adib Rastegarnia \IEEEauthorrefmark{2}}
\IEEEauthorblockA{ Dept of Computer Science,
Purdue University, USA, IN, 47907\\
Email: \IEEEauthorrefmark{1}comer@cs.purdue.edu,
\IEEEauthorrefmark{2}arastega@purdue.edu
}}


%


\maketitle

\begin{abstract}
Software Defined Networking (SDN) offers flexibility to program a network based on a set of network requirements. Programming the networks using SDN is not completely straightforward because a programmer must deal with low level details. To solve the problem, researchers proposed a set of network programming languages that provide a set of high level abstractions to hide low level hardware details. Most of the proposed languages provide abstractions related to packet processing and flows, and still require a programmer to specify low-level match-action fields to configure and monitor a network. Recently, in an attempt to raise the level at which programmers work, researchers have begun to investigate Intent-based, descriptive northbound interfaces. The work is still in early stages, and further investigation is required before intent-based systems will be adopted by enterprise networks.

To help achieve the goal of moving to an intent-based design, we propose an SDN-based network programming framework, the \emph{Open Software Defined Framework (OSDF)}. OSDF provides a high level Application Programming Interface (API) that can be used by managers and network administrators to express network requirements for applications and policies for multiple domains. OSDF also provides a set of high level network operation services that handle common network configuration, monitoring, and Quality of Service (QoS) provisioning. OSDF is equipped with a policy conflict management module to help a network administrator detect and resolve policy conflicts. The paper shows how OSDF can be used and explains application-based policies. Finally, the paper reports the results of both testbed measurements and simulations that are used to evaluate the framework from multiple perspectives, including functionality and performance.   
\end{abstract}

\section{Introduction}
In recent years, \emph{Software Defined Networking (SDN)} has received attention from both academia and industry for the design of network management systems. SDN provides the ability to program networks directly by breaking the vertical integration of control and data planes\cite{Ref1}. The separation of control and data planes allows network switches to become simple forwarding devices and gives us flexibility to control and program them by implementing the control logic in a centralized SDN controller using software called a \emph{Network Operating System (NOS)}\cite{Ref2}. In the current SDN paradigm, SDN controllers compromises three key layers including data plane, control plane, and application layers. Most of the SDN controllers employ two \emph{Application Programming Interfaces (APIs)} called the \emph{Northbound} and \emph{Southbound} APIs. Network applications use northbound APIs to communicate with the controller, express network behaviors, define configuration requirements, and program forwarding devices. Southbound APIs define the communication protocols between the controller and network devices.  The \emph{OpenFlow} protocol \cite{Ref4} defines a well-known and especially prominent southbound API, used by a SDN controller to update and insert flow rules that specify associated actions to be performed for each of the flows that pass through a given network device. Most of the SDN controllers provide a simplified northbound API, such as a RESTful API, that can be used to program network devices by specifying the details of flow rules using JSON or XML formats. With such an interface, a programmer must parse JSON or XML formats to retrieve required information (e.g., network topology, statistics, and parameters) which can then be used to program network devices. Although some SDN controllers provide high level services to make the network programming easier, a programmer still needs to deal with low level flow rule details. 

Although, there is no standard Northbound Interface (NBI) for SDN controllers \cite{8004109,intentNBI}, adopting an \emph{intent-based} interfaces for SDN seems promising. An Intent NBI is independent of a specific network technology, and uses application related vocabulary and information.  An Intent NBI provides abstractions that hide low level details of the network objects and services, and can be used by users to express their intents in a descriptive manner instead of a prescriptive manner.  In addition, an Intent NBI can be used to express expectations regarding the service controller will deliver.  Work has begun on intent based NBIs, including the ONOS intent framework \cite{onosintent} and NEMO \cite{nemo}.  However, further investigation of intent-based NBIs is warranted. In this paper, we attempt to answer some of the questions surrounding the intent-based approach, including:
\begin{itemize}
\item How a policy conflict management module be integrated with an intent based SDN programming framework to resolve network configuration conflicts at the intent level?
\item How can the flexibility of a reactive approach be incorporated into an intent based SDN programming framework?
\item How can an intent-based language be designed and implemented that allows managers to to express network requirements for application and policies for multiple domains?
\end{itemize}

This paper is an extension an earlier work-in-progress paper \cite{8319173}. The previous paper introduced the main modules of the framework and reported preliminary simulation results. Our main contributions in this paper are:
\begin{itemize}
\item Adds new network operations to the framework to support new services such as QoS provisioning and and an alerting mechanism to log user unauthorized attempts. 
\item Adds policy conflict management module to detect policy conflicts.
\item Describes an implementation of typical SDN applications that demonstrate our high-level policy language.
\item Adds experimental testbed measurements as well as additional simulations to assess the functionality and performance of the framework.
\end{itemize}

The essence of the work consists of a policy-based network programming framework called \emph{Open Software Defined Framework (OSDF)} that started with the following design goals:

\begin{itemize}
\item Provide a high level API that allows managers to express network configuration requirements using application based and domain specific network policies, and allows them to write management applications without worrying about low level details such as flow rule details, network topology related information (e.g end to end paths), etc. 
\item Define a set of high level network services that can
be invoked by management applications to configure network
switches and provide Quality of Service (QoS) without knowing the details of the southbound API (e.g., OpenFlow or an alternative).

\item Design and develop a framework that can run management applications
similar to the conventional operating systems which
run processes. A network management application uses services provided by
our framework similar to a conventional process that uses services provided by its operating system.
\item Devise a hybrid approach which allows programmers to specify network configuration requirements both proactively by deriving configuration flow rules from high level network policies and, reactively, by modifying flow rules as flows and conditions change.   

\item Design and develop a policy conflict management system to detect well known types of policy conflicts and recommend  potential conflict resolution solutions to the managers.  

\end{itemize}

The rest of the paper is organized as follows. Section \ref{Section 2} presents an overview of OSDF architecture and explains its key components in details.  Section \ref{policies} explains application based network policies that we use in our framework. Section \ref{hlnp} presents high level network operation services that our framework supports. Section \ref{pcds} introduces the Policy Conflict Management module and its key services. Section \ref{setup} explains the experimental and simulation environments used in the paper. Section \ref{results} presents our experimental and simulation scenarios and measurements. We present  the
future  work  in  Section \ref{futurework}. Section \ref{conclusion} concludes the paper.

\section{An Overview of OSDF Architecture}
\label{Section 2}
\begin{figure}[ht]
\centering
\includegraphics[scale=0.27]{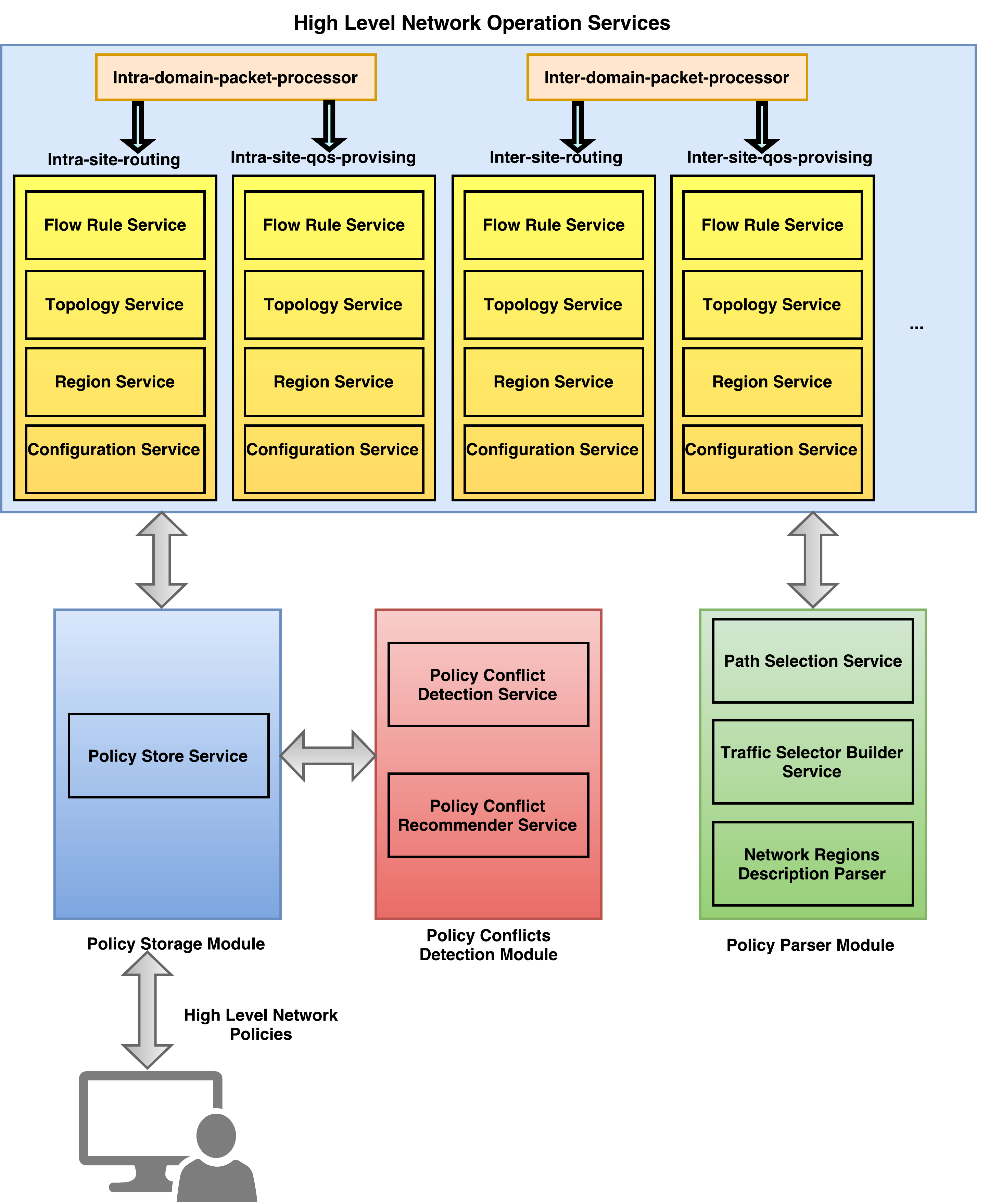}
\caption{OSDF architecture}
\label{Fig1}
\end{figure}

Fig.\ref{Fig1} illustrates the architecture of OSDF and explains its key components.

\begin{itemize}
\item \emph{High level network operation services}: To configure and monitor a network based on high level network policies that a network administrator provides, we provide a set of high level network operation services.  Each service takes the following steps to configure network switches based on current active policies in the system:
\begin{enumerate}
\item First, each service reads the high level network policies from a database of currently active policies and filters them based on the type of service. 
\item Second, each service uses the Policy Parser module to parse filtered policies and generate network-wide forwarding rules for incoming flows.
\item Finally, each service uses Flow Rule, Topology, Region and Configuration services to install generated rules in appropriate network switches. 
\end{enumerate}

Each service uses a hybrid approach to create basic rules proactively from the high level requirements that are specified in the network policies and to extract low level information reactively from incoming packets and use the results to generate, install, and update flow rules. 
Each network operation service includes the following subcomponents: 
\begin{itemize}
\item \emph{Packet Processor}: 
This component is responsible for parsing incoming packets reactively. That is, it reacts to packets that reach the controller by extracting low level match fields (e.g., IP addresses, MAC addresses, ports, and protocol number). Reactively processing packets allows a network administrator to defer configuration and hide low level details that are needed to configure and monitor a network. Packet processors parse the current active policies, which are stored in the policy database, and use \emph{Traffic Selector Builder Service} to generate a subset of match fields for a flow based on the application type and high level requirements for the type that an administrator specifies in the network policies. 
A Packet Processor service chooses an appropriate action for each flow rule, based on the high level operation. For example, in intra-site routing, the flow rules associated with a given flow specify forwarding packets to appropriate outgoing port in each switch.  For inter-site routing, rules may specify rewriting the MAC address, VLAN tagging, encapsulation, or other actions.  As Fig.\ref{Fig1} illustrates, we use two packet processors to provide intra-domain and inter-domain network operations, keeping the two logically separate them which simplifies the code.

\item \emph{Flow Rule Service}:  The Flow Rule Service is responsible for generating and installing OpenFlow rules in appropriate network switches.  It uses the set of match-action fields generated by the Traffic Selector Builder Service.

\item \emph{Topology Service}:  OSDF uses a Topology Service to find and determine an appropriate path for incoming flows based on high level network policies and the interconnections among network switches.  The \emph{Path Selection Service} uses the topology information that Topology Service provides to determine an end to end path according to requirements that an administrator provides in a network policy.

\item \emph{Region Service}: A \emph{region} refers to a group of devices located in a common physical (i.e., geographical) or logical region.   The Region Service provides information about devices inside a region.  The network regions description parser uses the information that regions service provide to distinguish intra and inter domain traffic flows.    

\item \emph{Configuration Service}: The Configuration Service provides an interface which can be used to access the items which are defined in a configuration file, including both details of individual devices, their IDs and locations, the IP prefixes used, the mapping of IP prefixes to regions, and predefined items, such as default gateways.    

\end{itemize}

\item \emph{Policy Store Module}: The Policy Store module stores and retrieves application-based network policies that an administrator enters to the system.  The module provides a policy store management service which is used by high level network operation services to read current active policies in the system. In addition, an administrator can list, update, and delete current network policies dynamically at runtime. 
\item \emph{Policy Parser Module}: The Policy Parser module is responsible to  analyze application-based policies and incoming flows and derive a set of match fields that are then used to generate a set of flow table rules. The module includes of the following three subcomponents, which are invoked by abstract operation services: 
\begin{itemize}
\item \emph{Path Selection Service}: This service is responsible to provide a set of pre-defined algorithms for choosing among a set of existing paths between two end points (e.g., shortest path).  A network programmer can extend this module by specifying additional path finding algorithms. 
\item \emph{Traffic Selector Builder}: This service generates a set of match fields based on application based policies and incoming packets.  A programmer can extend this module by defining new types of applications and specifying combinations of packet match fields for each application.

\item \emph{Network Regions Parser}: The Network Region Parser module uses information that the Region Service provides, and parses incoming flows and categorizes them based on the regions they span. 
\end{itemize}
\item \emph{Policy Conflicts Management Module}: This module is responsible for detecting well known conflicts between the current active policies and providing potential high-level solutions to the network administrator.  The module includes the following subcomponents:  
\begin{itemize}
\item \emph{Policy Conflicts Detection Service}: This service implements a conflict detection algorithm, as explained in subsection \ref{pcds1}.
\item \emph{Policy Conflicts Recommender Service}: This service implements a policy conflict resolution algorithm that suggests potential ways to resolve conflicts among existing policies. Subsection \ref{pcrrm} explains the operation in detail.   
\end{itemize}
\end{itemize}

\begin{figure}[ht]
\centering
\includegraphics[width=\columnwidth]{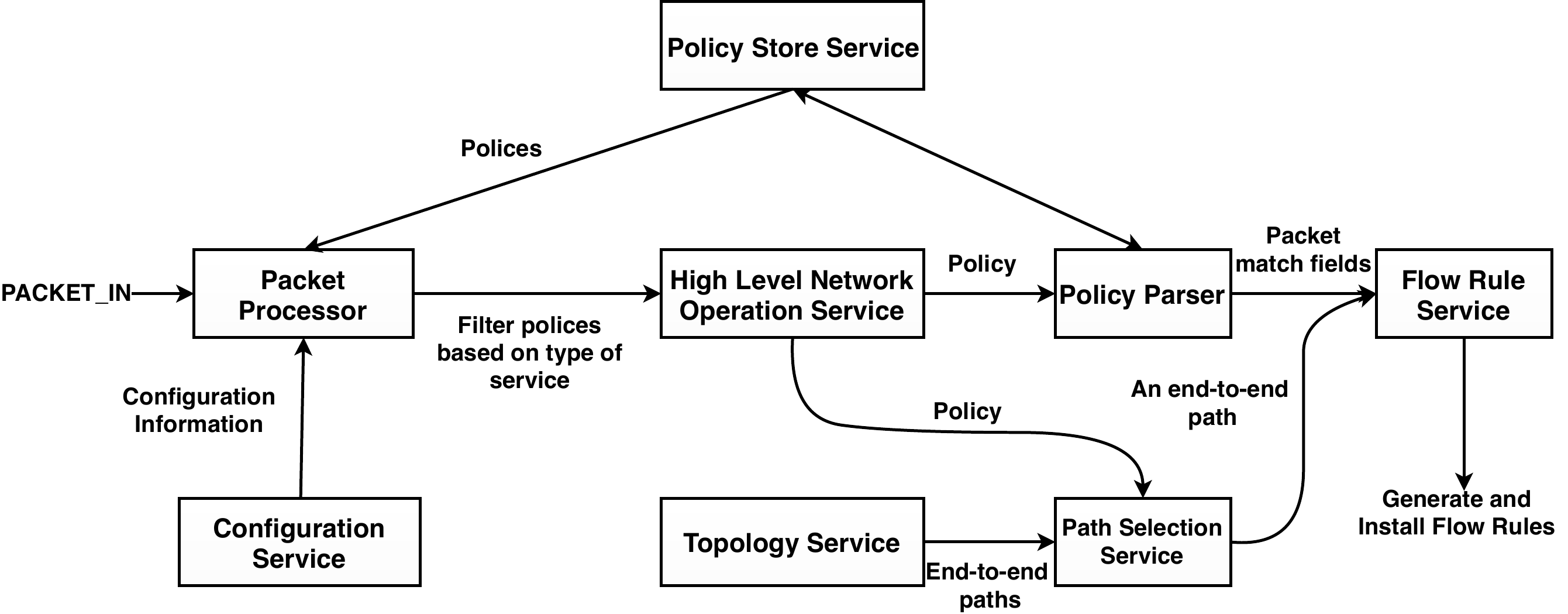}
\caption{OSDF packet processing and flow rule installation procedure}
\label{osdf_process}
\end{figure}

We illustrate the packet processing and flow rule installation steps in Fig.\ref{osdf_process}.

\section{Application-Based Network Policies}
\label{policies}
An application-based network policy specifies high level network requirements for a given application (type of packet).
The policies are used to configure and monitor network devices.  We divide all policies into two major categories: \emph{intra-domain policies} (inside a region or a site) and \emph{inter-domain policies} (among multiple regions or sites).  An application policy includes the following key items:
\begin{itemize}
\item \emph{Traffic Profile}: A traffic profile that specifies high level characteristics and requirements for an application, such as an application name (e.g. WEB), the transport protocol used (e.g TCP or UDP), and a traffic type (e.g. real time vs best effort). The system provides a set of pre-defined traffic profiles that support typical uses, such as web, video, and voice traffic. A network administrator can extend traffic profiles by introducing new traffic types and applications.
\item \emph{High Level Network Function}: Each policy is associated with a high level network function that defines configuration and monitoring of network devices, such as intra-site-routing and inter-site-routing. Associating each policy with a network operation allows a Packet Processor to accommodate policies that are related to a specific function and ignore non-relevant functions.   

\item \emph{Partial Hosts And Devices Information (address space conditions)}: An administrator has the flexibility to provide high-level information about devices and hosts (e.g specify a name for a host or a network device). Path Selection Service uses these high level information to determine an end-to-end path between the source and destination for a specific traffic. For example, an administrator can use device names when specifying that a given type of traffic should pass through a specific set of network devices; the Path Selection Service will choose the best possible path that meets the given requirements. The system uses partial host information when reporting network policy conflicts.  

\item \emph{Traffic Conditions}: Traffic conditions specify high level QoS requirements such as the traffic rate limit for a traffic which specified in the policy. 

\item \emph{Priority}: An administrator can assign a priority to a policy or use a default.  Priorities become important when the systems tries to resolve conflicts among policies.

\item \emph{Source And Destination Regions}: A policy can be defined for the interior of a region or can specify traffic routing among multiple regions. To achieve the goal, the system allows a manager to specify both source and destination regions for each policy.   
\end{itemize}

\begin{figure}[ht]
\centering
\includegraphics[width=\columnwidth]{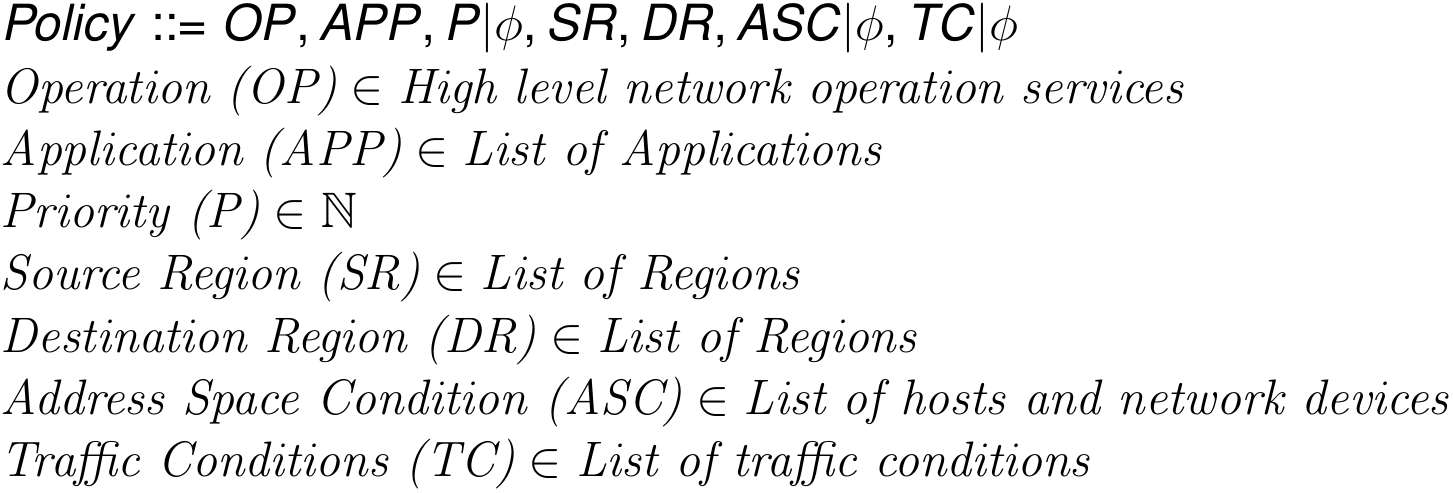}
\caption{The syntax of OSDF policy language}
\label{FigPolicy}
\end{figure}

The syntax of OSDF policy language is defined in Fig.\ref{FigPolicy}.

\section{High level Network Operations}
\label{hlnp}
In the initial prototype version of the framework we define a minimum set of high level network operations to support typical network configurations:

\begin{itemize}
\item \emph{Intra-Site-Route:} This abstract operation can be used to specify traffic routing within a specific region according to a network policy for the region.
\item \emph{Inter-Site-Route:} This abstract operation can be used to specify traffic routing among multiple regions according to the global policies.
\item \emph{Intra-Site-Alert:} This abstract operation can be used to set an alert on specific data traffic and users. If a user attempts to send the traffic inside a region, the system logs user unauthorized attempts and avoid a specific traffic to be routed inside a region.     
\item \emph{Inter-Site-Alert:} This abstract operation is the same as Intra-Site-Alert except it applies the same operation between multiple regions. 

\item \emph{Intra-Site-Route-QoS-Provisioning}: This abstract operation can be used for simultaneous traffic routing and rate limiting within a specific region according to a network policy for the region.

\item \emph{Inter-Site-Route-QoS-Provisioning}: This abstract operation can be used for simultaneous traffic routing and rate limiting among multiple regions according to the global policies. 

\end{itemize}

\section{Policy Conflict Management Module}
\label{pcds}
\subsection{Policy Conflict Detection Service}
\label{pcds1}

In \cite{Ref14}, the authors classify flow rule conflicts based on flow rule details such as layer 2-4 addresses, action, and priority. We use their work as a reference to define types of network policy conflicts based on high level application based network policies instead of low level flow rule details. As we illustrate in Table \ref{tbl:1} , we categorize some of well known network policy conflicts based on application based network policies and explain them as follows:

Suppose two policies, ${P_i}$ and ${P_j}$, specify the same traffic profile, source and destination regions:  

\begin{itemize}
\item \emph{Redundancy}: ${P_i}$ is redundant to ${P_j}$ if both specify the same network operation (${P_{i,OP}} = {P_{j,OP}}$), address space condition of ${P_i}$ is a subset of  address space condition of ${P_j}$ (${P_{i,SC}} \subseteq {P_{j,SC}}$), and priority of  ${P_i}$ is less than or equal to priority of ${P_j}$ (${P_{i,p}} \le {P_{j,p}}$). For example, suppose the following two policies:
\begin{enumerate}
\item Route web traffic inside region $A$ using priority 100 between hosts $H1$ and $H2$.
\item Route web traffic inside region $A$ using priority 100 between all hosts.
\end{enumerate}
In the example, the first policy is redundant because second policy is broader (assuming that H1 and H2 are in region $A$). 
\item \emph{Shadowing}:  ${P_i}$ is shadowed ${P_j}$ if each policy specifies a different network operation (${P_{i,OP}} \ne {P_{j,OP}}$), the address space condition of ${P_i}$ is a subset of the address space condition of ${P_j}$ (${P_{i,SC}} \subseteq {P_{j,SC}}$), and the priority of ${P_i}$ is less than to priority of ${P_j}$ (${P_{i,p}} < {P_{j,p}}$). Shadowing is a critical error because it shows a conflict in a security policy implementation \cite{Ref15}. For example, suppose the following two policies:
\begin{enumerate}
\item Route VIDEO traffic between all hosts in region $A$ and all hosts in region $B$ using priority 100.
\item Set alert on VIDEO traffic between any hosts in region $A$ and region $B$ using priority 200.
\end{enumerate}
The first policy is shadowed by the second policy and will never be invoked because its priority is less than the priority of second policy. 

\item \emph{Generalization}: ${P_i}$ is a generalization of ${P_j}$ if each policy specifies a different network operation (${P_{i,OP}} \ne {P_{j,OP}}$), address space condition of ${P_i}$ is a superset of  address space condition of ${P_j}$ (${P_{i,SC}} \supseteq {P_{j,SC}}$), and priority of  ${P_i}$ is less than  to priority of ${P_j}$ (${P_{i,p}} < {P_{j,p}}$).  For example, suppose the following two policies:
\begin{enumerate}
\item Route VOICE traffic inside region $A$ using priority 200 and all host pairs.
\item Set alert on VOICE traffic inside region $A$ using priority 300 and host pair (H3,H4). 
\end{enumerate}
The first policy is a generalization of the second policy because its address space is a superset of address space of the second policy (assuming that H3 and H4 are in region $A$). 

\item \emph{Correlation}: ${P_i}$ is a correlation of ${P_j}$ if each policy specifies a different network operation (${P_{i,OP}} \ne {P_{j,OP}}$), address space condition of ${P_i}$ is not a subset or superset of  address space condition of ${P_j}$ but they have some common items (${P_{i,SC}} \cap {P_{j,SC}} \ne \phi$), and priority of  ${P_i}$ is less than  to priority of ${P_j}$ (${P_{i,p}} < {P_{j,p}}$).  For example, suppose the following two policies:
\begin{enumerate}
\item  Route WEB traffic inside region $A$ using priority 100 and host pairs (H1,H2), (H1,H3).
\item  Set alert on WEB traffic inside region $A$ using priority of 200 and host pairs (H1,H2), (H2,H4).
\end{enumerate}
The address space of first policy is not a subset or superset of address space of the second policy but their intersection is not empty. Consequently, there is a correlation conflict between above policies (assuming that H1,H2,H3, and H4 are in region $A$).   
\item \emph{Overlap}: ${P_i}$ is an overlap of ${P_j}$ if each policy specifies the same network operation (${P_{i,OP}} = {P_{j,OP}}$), address space condition of ${P_i}$ is not a subset or superset of  address space condition of ${P_j}$ but they have some common items (${P_{i,SC}} \cap {P_{j,SC}} \ne \phi$). Note that priority is not important in this case.  For example, suppose the following two policies:
\begin{enumerate}
\item  Route WEB traffic between regions $A$ and region $B$ using priority 100 and host pairs (H1,H4), (H2,H3).
\item  Route WEB traffic between region $A$ and region $B$ using priority 200 and host pairs (H1,H4), (H2,H5).
\end{enumerate}
Both of the policies specify the same network operation and the intersection of the first policy address space and second policy address space is not empty. Consequently, there is an overlap conflict between above policies (assuming that H1,H2 are in region $A$ and H3,H4, and H5 are in region $B$). 
\end{itemize}

\begin{table*}[ht]
  \centering

  \begin{tabular}{|l|l|l|l|l|l|l|}
  \hline
    & Traffic Profile (TP) & Src Region (SR)    & Dst Region (DR) & Operation (OP) & Address Space Conditions (SC) & Priority (P) \\ \hline
    Redundancy & ${P_{i,TP}} = {P_{j,TP}}$ & ${P_{i,SR}} = {P_{j,SR}}$ & ${P_{i,DR}} = {P_{j,DR}}$ & ${P_{i,OP}} = {P_{j,OP}}$ & ${P_{i,SC}} \subseteq {P_{j,SC}}$ & ${P_{i,p}} \le {P_{j,p}}$ \\ \hline
  
        Shadowing & ${P_{i,TP}} = {P_{j,TP}}$ & ${P_{i,SR}} = {P_{j,SR}}$ & ${P_{i,DR}} = {P_{j,DR}}$ & ${P_{i,OP}} \ne {P_{j,OP}}$ & ${P_{i,SC}} \subseteq {P_{j,SC}}$ & ${P_{i,p}} < {P_{j,p}}$ \\ \hline
               Generalization & ${P_{i,TP}} = {P_{j,TP}}$ & ${P_{i,SR}} = {P_{j,SR}}$ & ${P_{i,DR}} = {P_{j,DR}}$ & ${P_{i,OP}} \ne {P_{j,OP}}$ & ${P_{i,SC}} \supseteq {P_{j,SC}}$ & ${P_{i,p}} < {P_{j,p}}$ \\ 
    \hline
           Correlation & ${P_{i,TP}} = {P_{j,TP}}$ & ${P_{i,SR}} = {P_{j,SR}}$ & ${P_{i,DR}} = {P_{j,DR}}$ & ${P_{i,OP}} \ne {P_{j,OP}}$ & ${P_{i,SC}} \cap {P_{j,SC}} \ne \phi $ & ${P_{i,p}} \le {P_{j,p}}$ \\ 
    \hline
     Overlap & ${P_{i,TP}} = {P_{j,TP}}$ & ${P_{i,SR}} = {P_{j,SR}}$ & ${P_{i,DR}} = {P_{j,DR}}$ & ${P_{i,OP}} = {P_{j,OP}}$ & ${P_{i,SC}} \cap {P_{j,SC}} \ne \phi $ & Not Necessary \\ 
    \hline
  \end{tabular}

  \caption{Different types of policy conflicts}
      \label{tbl:1}
\end{table*}

\subsection{Policy Conflict Recommendation Service}
\label{pcrrm}
This service implements a policy conflict recommendation algorithm  to provide high level advice about how an administrator can resolve the conflicts between current policies in the system. Each type of conflict is resolved as follows:
\begin{itemize}
\item \emph{Redundancy}: To resolve this type of conflict, we can remove a policy that has a lower or equal priority and its address space that is a subset of another policy.  
\item \emph{Shadowing}: To resolve this type of conflict between two policies, we can remove the policy that is shadowed (i.e. the policy which has lower priority and its address space is a subset of another policy).
\item \emph{Generalization}: To resolve this type of conflict between two policies, we can remove the policy that has a more generalized address space, and then update its address space conditions by removing those conditions that are specified in another policy address space. Finally, we insert the updated policy into the system. 
\item \emph{Correlation}: To resolve this type of conflict between two policies, we can update the address space of one of the policies (e.g the policy by lower priority) by removing the common items and then inserting the updated policy into the system. 
\item \emph{Overlap}: To resolve this type of conflict between two policies, we first remove both of the policies from the system and then insert a new policy with an address space equal to the union of both of address spaces. 
\end{itemize}
The asymptotic time complexity of the policy conflict recommendation algorithm is $O(n^2)$ where n is the number of polices being considered.  Most practice, n is small because a single policy covers a broad range of network configurations.

\section{Prototype and Experimental Setup}
\label{setup}
An early version of OSDF is implemented in \emph{Open Network Operating System (ONOS)} \cite{ Ref17}, an open source SDN controller which like other SDN controllers, provides services and APIs that programmers use to write applications and modules inside ONOS. We use ONOS APIs, such as the topology and flow rule APIs, to implement each network operation. Other modules, such as the policy parser, policy store, and policy conflict detection modules, have been implemented from scratch. To evaluate OSDF from multiple perspectives such as functionality and performance, we use both experimental measurement and simulation. For experimental results, we deployed an SDN testbed which we explain its details as follows:  

\subsection{SDN Experimental Testbed}
\begin{figure}[ht]
\centering
\includegraphics[scale=0.21]{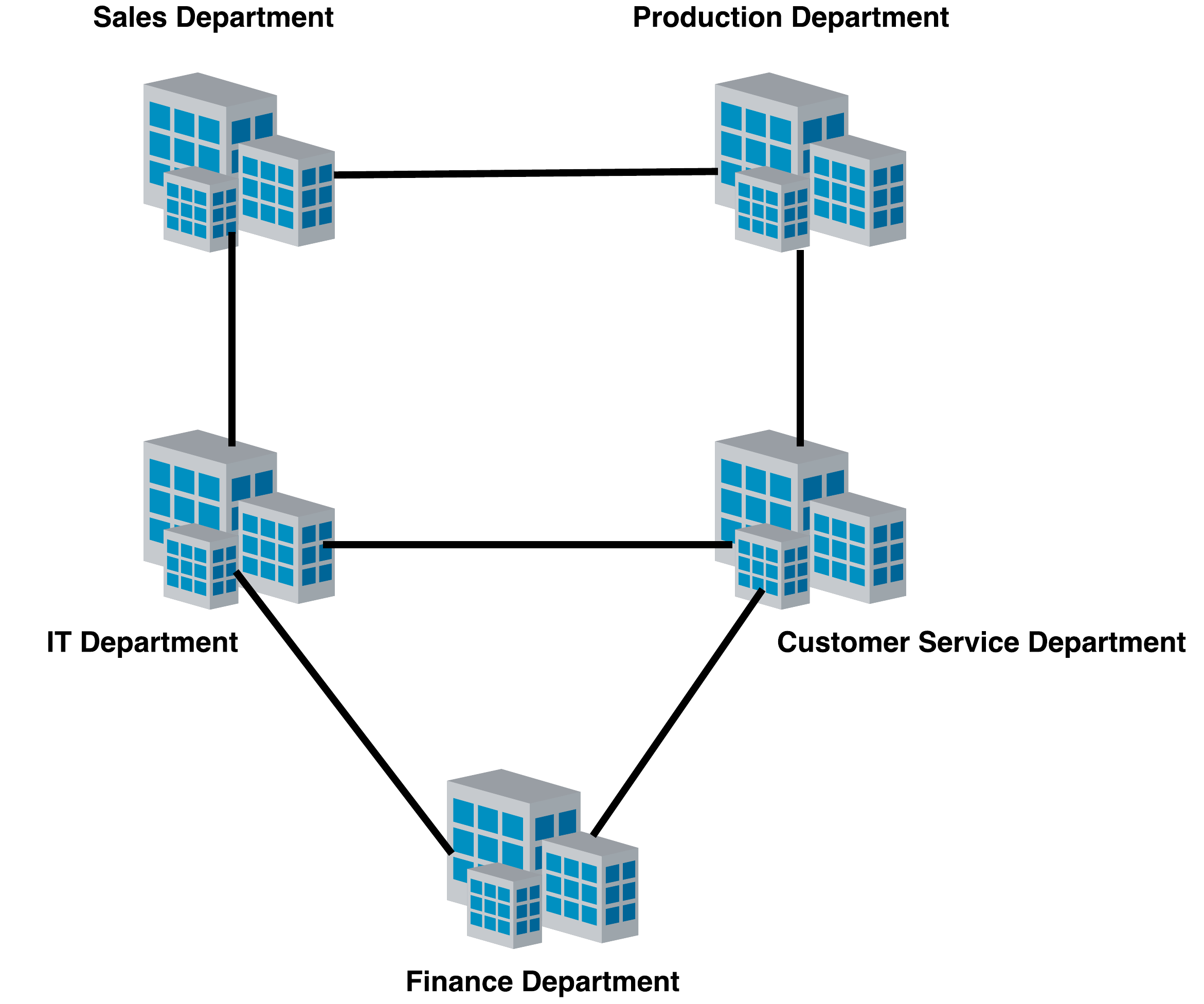}
\caption{Inter-sites network topology}
\label{Fig2}
\end{figure}

Our SDN testbed consists of 10 Ethernet switches that logically define 5 sites (departments of an organization) that are connected together as Fig.\ref{Fig2} illustrates. To make our testbed emulate an enterprise network, each physical switch is logically divided into 10 independent smaller switches using virtualized mode. Each site is a Fat-tree network topology. The connections between virtual switches are $1$GB links and we aggregate $1$GB links to connect sites, thereby emulating high capacity links. Overall, our network consists of $50$ virtual switches and more than $130$ links.

\subsection{Simulation Setup}
In addition to an SDN testbed, we used the \emph{Mininet} \cite{Ref18} network emulator to evaluate our framework at a larger scale.  We ran Mininet on a Linux host with a Intel Core i7 processor and 8GB RAM.

\subsection{Traffic Generator}
To generate various traffic patterns, we used \emph{iperf} \cite{Ref28}, a tool for measuring throughput on IP networks. 

\section{Simulation and Experimental Results}
\label{results}

 This section presents typical applications such as \emph{Network Traffic Isolation}, \emph{Inter Domain Rate Limiter} to provide QoS,  \emph{Inbound Traffic Engineering}, and \emph{Forwarding Resiliency}; OSDF allows each to be implemented easily.

\begin{enumerate}
\item \emph{\textbf{Traffic Isolation}}:
Consider the topology a leaf spine network topology (we call it site A in this example) that Fig.\ref{Fig9} illustrates.

\begin{figure}[ht]
\centering
\includegraphics[scale=0.25]{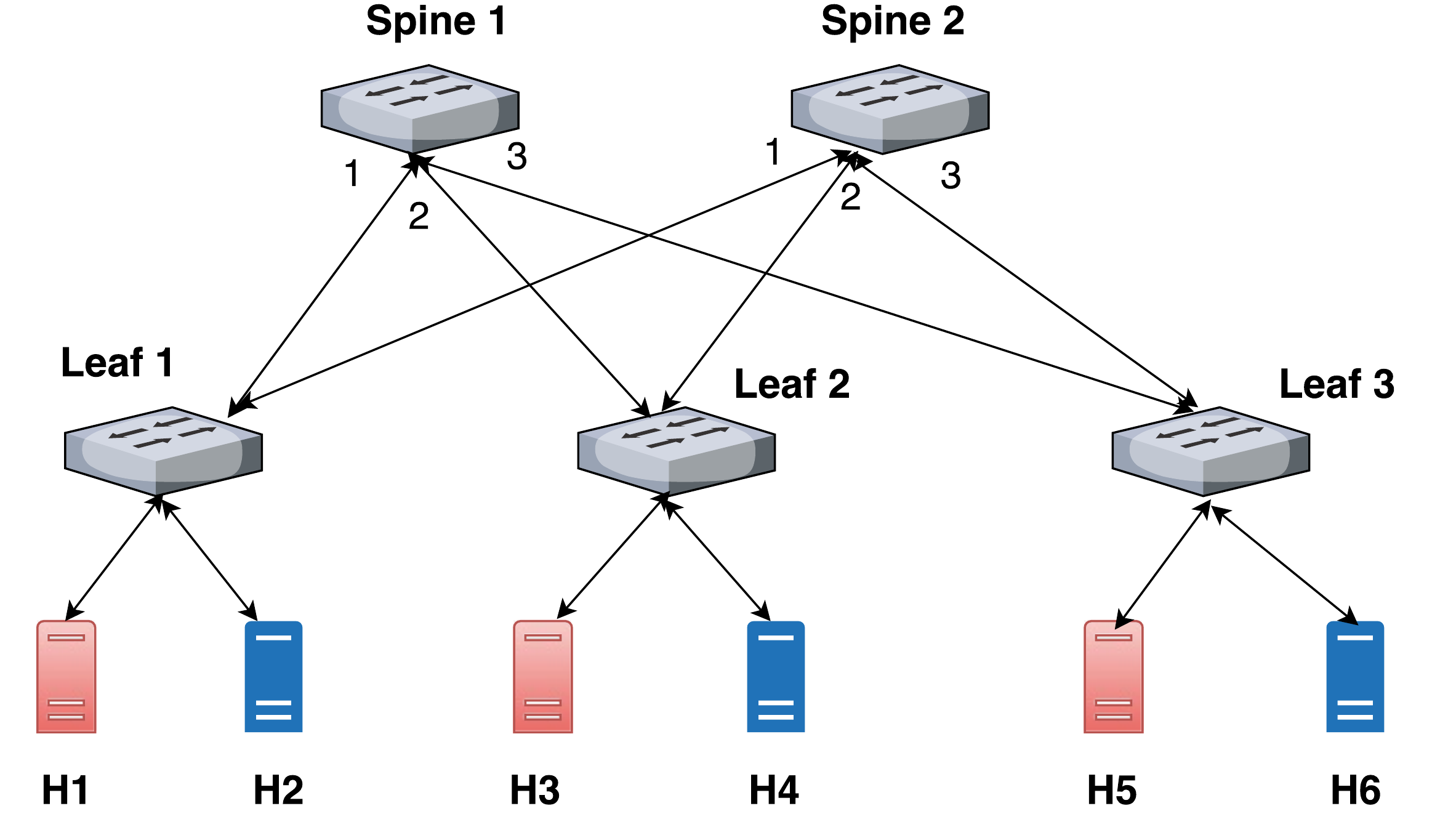}
\caption{Leaf and Spine network topology for traffic isolation scenario}
\label{Fig9}
\end{figure}
Suppose H1, H3, and H5 belong to tenant 1, and H2,H4, and H6 belong to tenant 2. We configure the network using OSDF such that hosts owned by a given tenant are able to communicate with each other but are blocked from communicating with hosts owned by other tenants. Suppose we want to route web traffic for tenant 1, and route video streaming traffic for tenant 2. In addition, suppose all of the network switches in the given topology belong to the same region, which we will label site $A$. To achieve the goal, we define OSDF network policies as follows:

\begin{itemize}
\item Route web traffic in site A between (H1, H3), (H1, H5), and (H3,H5) using the default priority.
\item Route video traffic in site A between (H2, H4), (H2, H6), and (H4,H6) using the default priority. 
\end{itemize}

The network administrator enters above policies into the system using a command line interface or a script, and OSDF installs OpenFlow rules in appropriate network switches to meet the requirements. As the example shows, instead of requiring managers to install flow rules, OSDF provides traffic isolation at the policy level. 

\item \emph{\textbf{Inter Site QoS Provisioning:}}
The main goal of this scenario is to experiment with using the abstractions defined above to provision QoS for inter site routing. OpenFlow 1.3 introduces meters which can be used to measure and control the ingress rate of traffic.  If the packet rate or byte rate passing through the meter exceed a predefined threshold, a meter band will be triggered. A \emph{Rate Limiter} is a meter which its meter band drops the packet. To achieve the goal, the underlying hardware must support metering.   Because Open vSwitch \cite{Ref27} does not support metering completely, we decided to test our QoS provisioning abstractions using our SDN testbed.  Suppose we want to configure the network illustrated in Fig.\ref{Fig2} to support the following QoS requirements:

\begin{itemize}
 
\item Route web traffic between IT and the sales department using the default priority, and limit each flow to  200 Mbps.
\item Route video traffic between IT and the sales department using the default priority, and limit each flow to 500 Mbps.
\end{itemize}

 We ran an experiment between IT and sales department to test inter-site routing with rate limiting. In this scenario, we initiated four TCP connections (an aggregate of 800 Mbps) to generate web traffic and two TCP connections (an aggregate of 1000 Mbps) to generate video traffic.  Fig.\ref{Fig6} summarizes the results, and shows that the throughput for each of the web traffic and video traffic flows is guaranteed to 200 Mbps and 500 Mbps, respectively.

\begin{figure}[ht]
\centering
\includegraphics[scale=0.63]{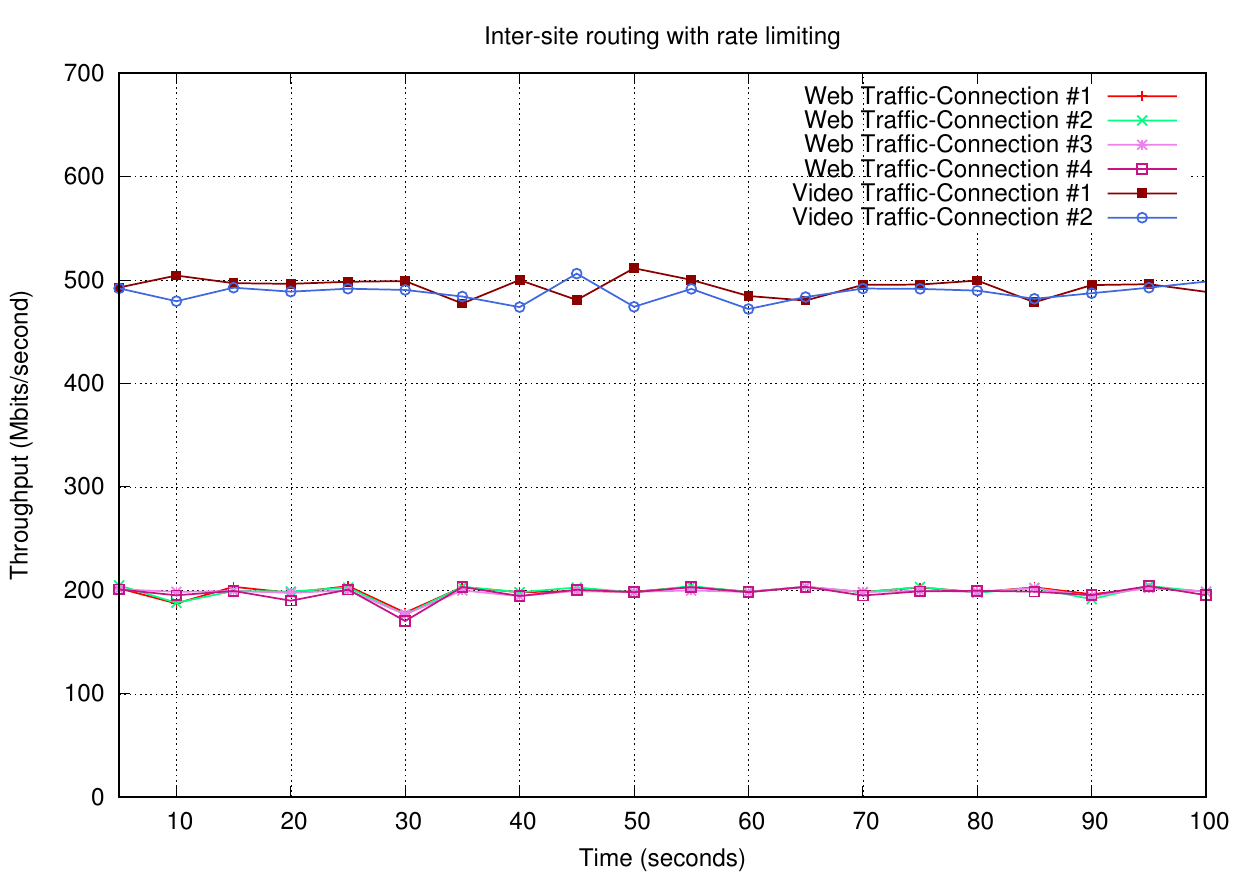}
\caption{Inter-site routing with rate limiting}
\label{Fig6}
\end{figure}

To examine how well QoS provisioning works, we measured inter-site routing without QoS.  To do so, we entered the same basic network policies for web and video traffic, omitting  the rate limiting policies:

\begin{itemize}
\item Route web traffic between IT and sales departments using the default priority.
\item Route video traffic between IT and sales departments using the default priority.
\end{itemize}

As the results in Fig.\ref{Fig8} show,  without rate limiting, video and web traffic are not guaranteed specific limits, and the throughput fluctuates more than when QoS is applied (i.e., more than in  Fig.\ref{Fig6}). 

Another proposed programming language, Merlin \cite{Ref35}, addresses QoS provisioning by providing high level abstractions that can be used to express bandwidth constraints such as minimum and maximum bandwidth limits. Compared to our approach, Merlin shares the same weakness as other network programming languages (e.g., NetKAT) because a programmer must specify low-level packet match fields to define a policy.

\begin{figure}[ht]
\centering
\includegraphics[scale=0.63]{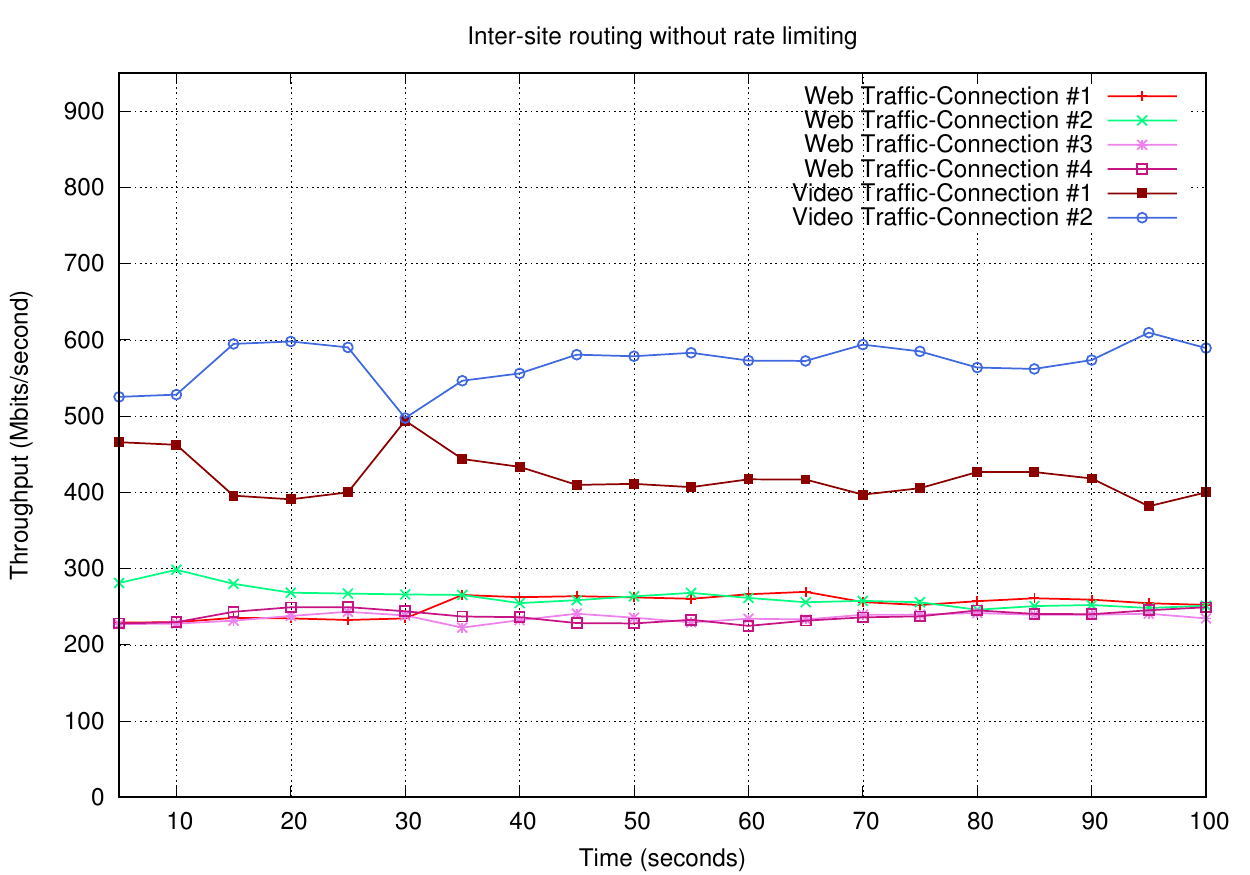}
\caption{Inter-site routing without rate limiting}
\label{Fig8}
\end{figure}

\item \emph{\textbf{Inbound Traffic Engineering}}: This service refers to splitting incoming traffic over multiple peering links. Consider, for example, the internal representation of the IT and Sales departments in our SDN testbed, as illustrated in Fig.\ref{Fig81}. Suppose we want to configure the two sites such that the incoming traffic to switch \emph{Sales-S1} should be split among the outgoing links. That is, web traffic from the IT department should be forwarded to port 2 of switch Sales-S1, and video streaming traffic should be forwarded to port 3 of the same switch.

\begin{figure}[t]
\centering
\includegraphics[width=\columnwidth]{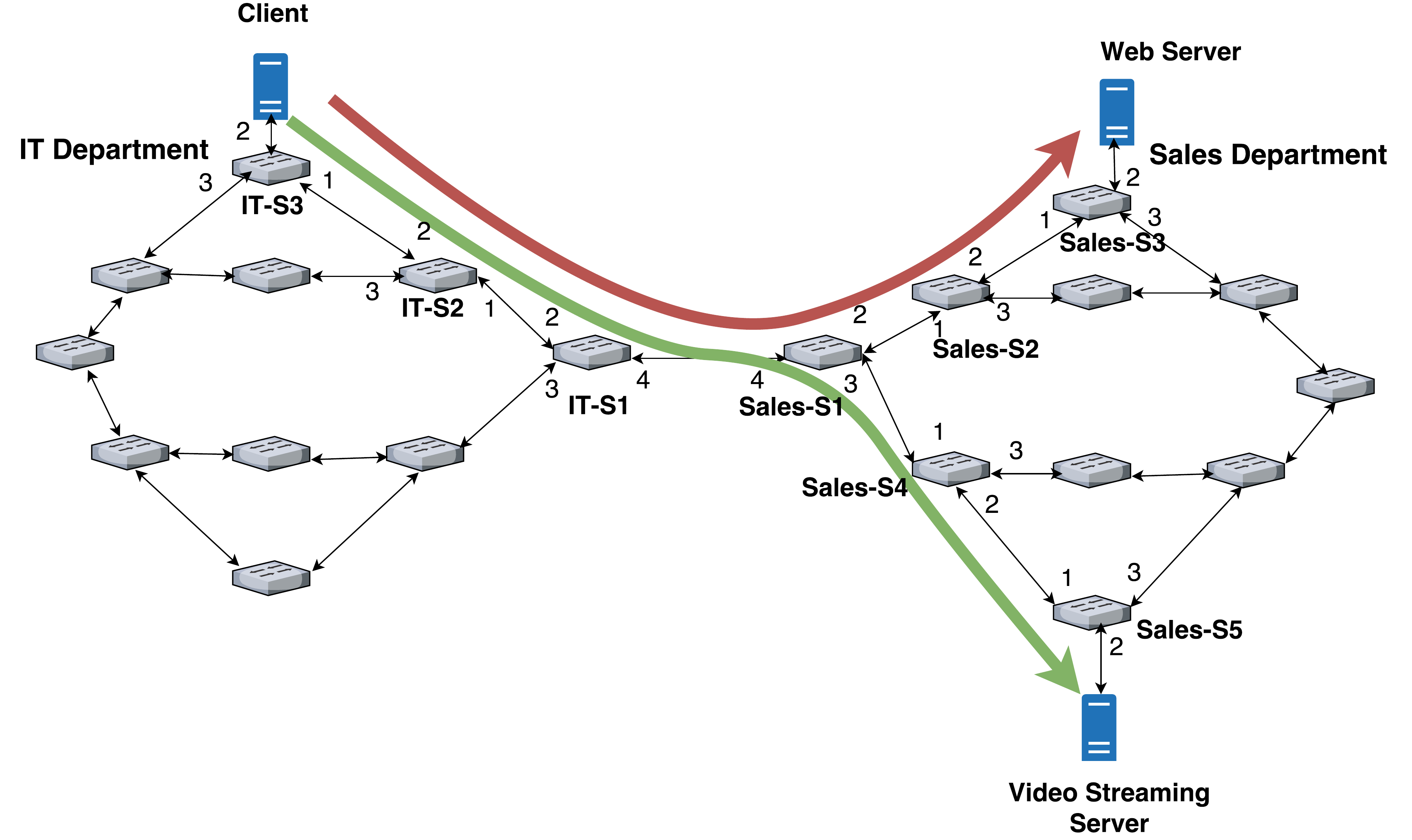}
\caption{Inbound traffic engineering scenario}
\label{Fig81}
\end{figure}

To achieve the goal, we can define high level policies as follows:
\begin{itemize}
\item Route web traffic between IT and Sales departments via Sales-S1:2 using the default priority.
\item Route video traffic between IT and Sales departments via Sales-S1:3 using the default priority.
\end{itemize}

\item \emph{\textbf{Forwarding Resiliency}}: Consider the network topology that Fig.\ref{Fig11} illustrates; call it site C.  We report an experiment that shows how OSDF can use redundant policies with differing priorities to handle the case of reroute traffic (web traffic in the example) from a primary path to a backup path during failures or for maintenance purposes. 
\begin{figure}[t]
\centering
\includegraphics[width=\columnwidth]{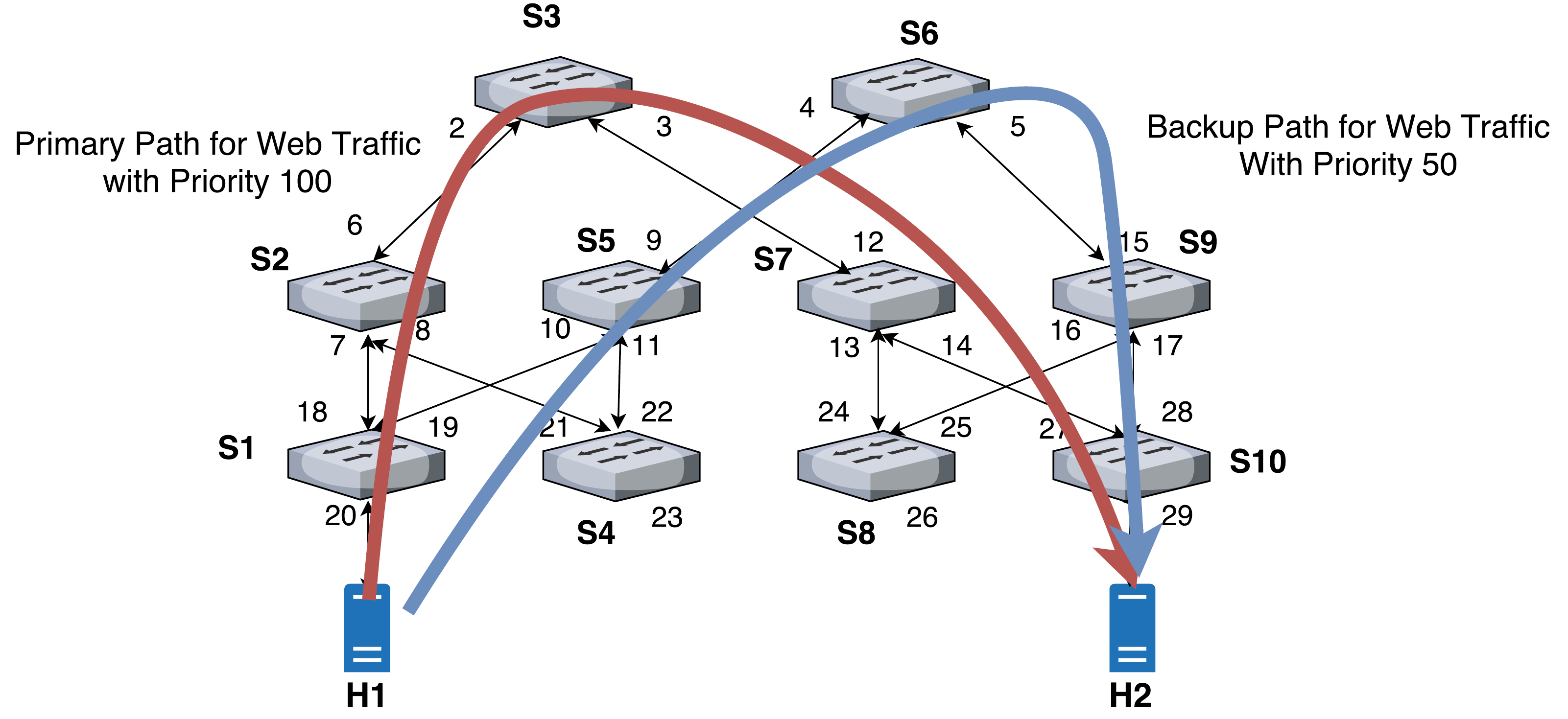}
\caption{A multipath network topology (site C)}
\label{Fig11}
\end{figure}
In the first step, we define high level policies for both a primary path and a backup path between two points:

\begin{itemize}
\item Route web traffic in site C using priority 100 via S3:2.
\item Route web traffic in site C using priority 50 via S6:4.
\end{itemize}

Based on the above policies, if host H1 initiates web traffic to host H2, the policy with higher priority will be triggered, and OpenFlow rules with priority 100 will be added to network switches along the primary path (i.e.  H1, S1, S2, S3, S7, S10, H2).\\ 
To show how OSDF reacts to adding and removing policies, we wrote a script that forces a change from the primary to a backup path by removing and adding the policies every $N$ (Interval) seconds in a 150 seconds simulation. One aspect of any networking system arises from the need to re-establish valid routes after a failure with minimal interference.  To demonstrate that our framework supports rerouting without negative impact on traffic, we conducted an experiment to compare the throughput of a flow in the presence of rerouting to the throughput of the same flow with no rerouting. To conduct the experiment, we ran iperf as a server using port 80 on H2, and ran iperf as a client on H1 with default Linux configuration. We repeated the experiment for various numbers of parallel connections.  As the results in Fig.\ref{Fig11.1} show, when switching between primary path and backup path is not frequent (e.g Interval=40) the throughput remains close to the maximum possible throughput, which is around 940 Mbps (achieved when just the primary path is used). When we change between the primary and backup paths frequently (e.g., every 2 seconds) the the throughput drops significantly, specially when multiple, parallel TCP connections (e.g., 50).  Multiple factors explain the decrease.  First, as the number of connections increases from 1 to 50, the setup time (i.e., time required to install OpenFlow rules) will increase because the the time is linearly proportional to number of rules. Second, switching paths introduces delay, which drives TCP into congestion avoidance.  Frequent changes means lower throughput because TCP will spend more time recovering, and less time in a stable state. 

\begin{figure}[t]
\centering
\includegraphics[scale=0.58]{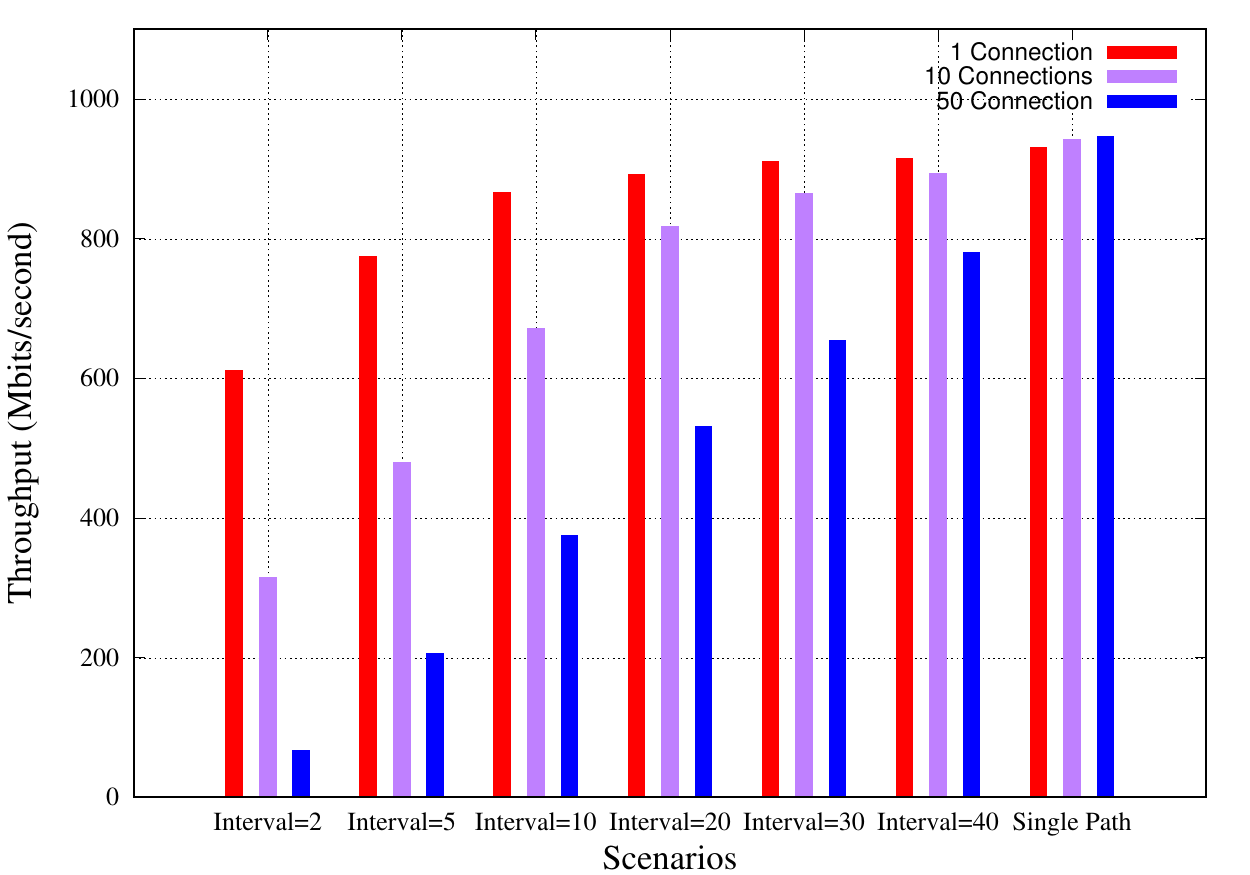}
\caption{Experimental results for forwarding resiliency scenarios}
\label{Fig11.1}
\end{figure}

\end{enumerate}

\subsection{Simulation Scenarios}
To test our framework from functional and performance points of view, and to examine the effects of scale, we ran simulation scenarios. 

\begin{itemize}
\item \emph{\textbf{Response time}}: The simulation uses a worst-case linear network topology with 40 switches that we call site D.  Fig.\ref{Fig14} illustrates the topology.  To configure the network to route web traffic between H1 and H40, we use a high-level policy:
\begin{itemize}
\item Route web traffic in site D using the default priority.
\end{itemize}

\begin{figure}[t]
\centering
\includegraphics[scale=0.32]{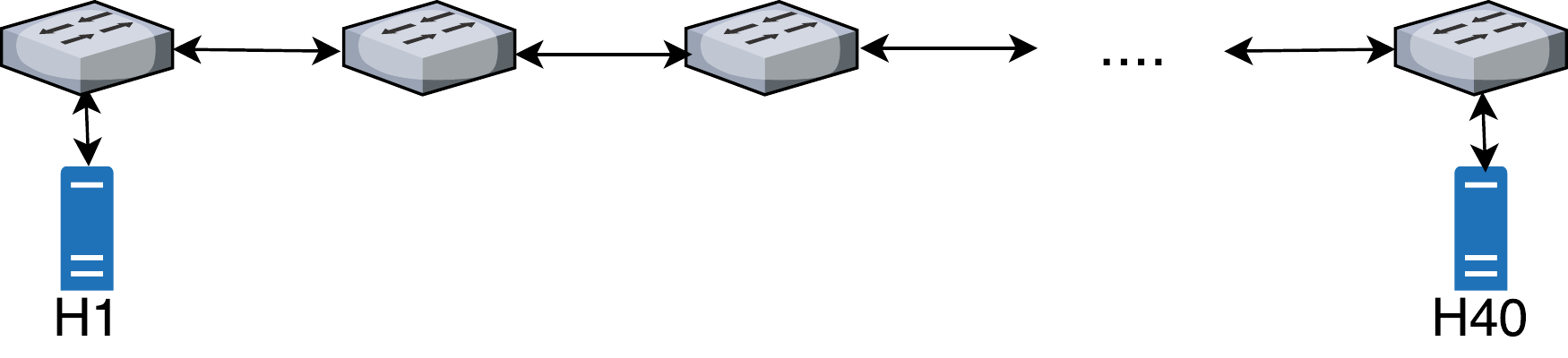}
\caption{Linear network topology}
\label{Fig14}
\end{figure}
To evaluate setup time, we initiate various numbers of TCP connections on port 80 between H1 and H40, and measure the response time for each. We define the response time as the amount of time which is needed to establish all of the connections between the two end-points. This is the amount of time which is needed to read and parse a network policy, generate flow rules , and install generated rules on the network devices for all of the connections between the two end-points. 
 By increasing the number of TCP connections established between two end-points, we increase the number of OpenFlow rules that must be installed in network switches along the path.  Consequently, we measure the response time as a function of number of flows and compared the response time of OSDF with reactive forwarding approach implemented by ONOS, and compare the two in Fig.\ref{Fig15}. As the results show, the response time increases linearly as number of connections increase between two end-points. The reason that our system outperforms the ONOS reactive approach arises from an optimization in which our system pre-installs OpenFlow rules in all switches across the entire end-to-end path when the first \emph{PACKET\_IN} message arrives at a controller.  The
ONOS reactive forwarding approach waits until a \emph{PACKET\_IN} message arrives from a switch before installing a forwarding rule in the switch.

To validate the simulation results, we ran the same experiment on our testbed; Fig.\ref{Fig16} shows the results.
The experimental measurement confirms that the simulation results are valid.  Note that the TCAM table on network switches in our testbed can hold up to 2K entries when more than 2 match fields are used by flow rules, so the number of connections is limited.

\begin{figure}[t]
\centering
\includegraphics[scale=0.59]{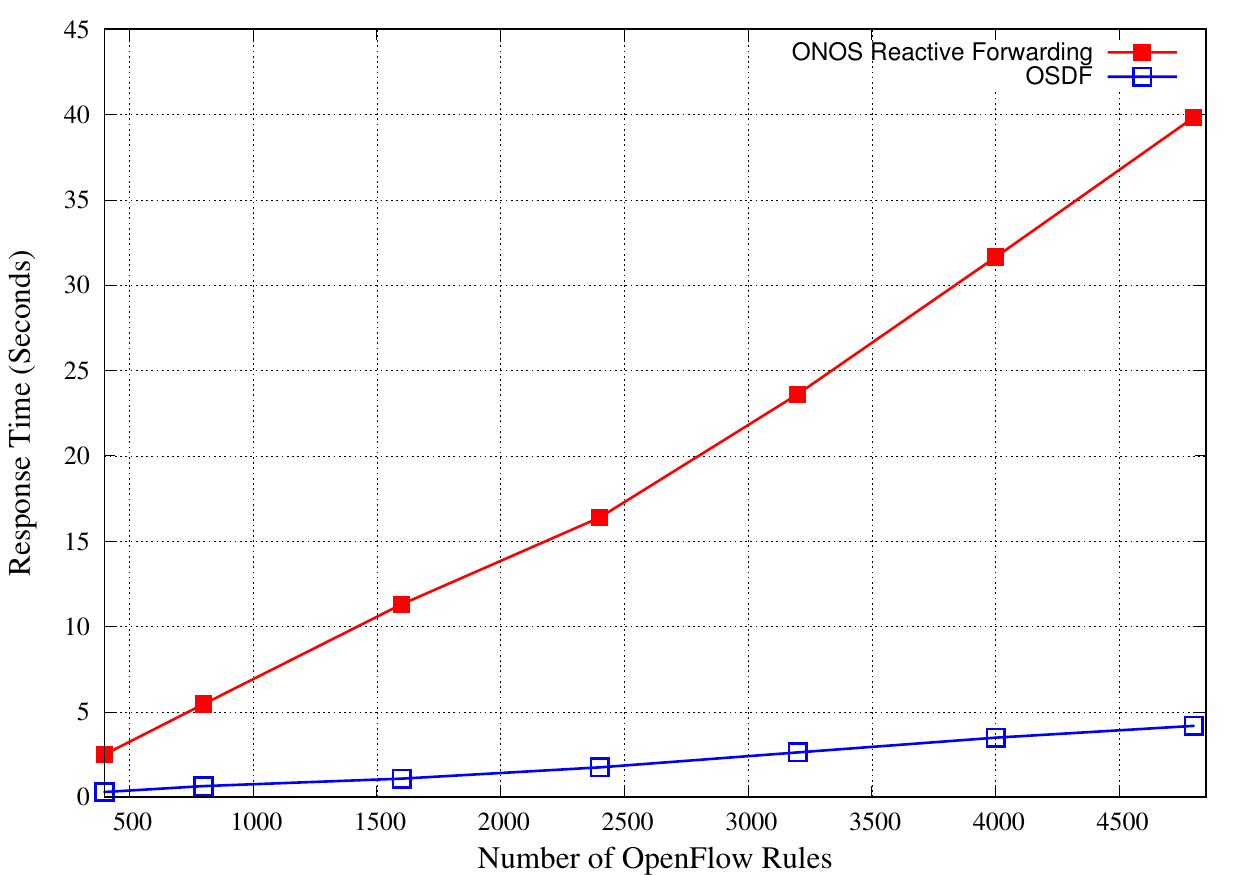}
\caption{Response time, ONOS reactive forwarding vs OSDF for a linear network topology (The simulation scenario)}
\label{Fig15}
\end{figure}

\end{itemize}

\begin{figure}[t]
\centering
\includegraphics[scale=0.59]{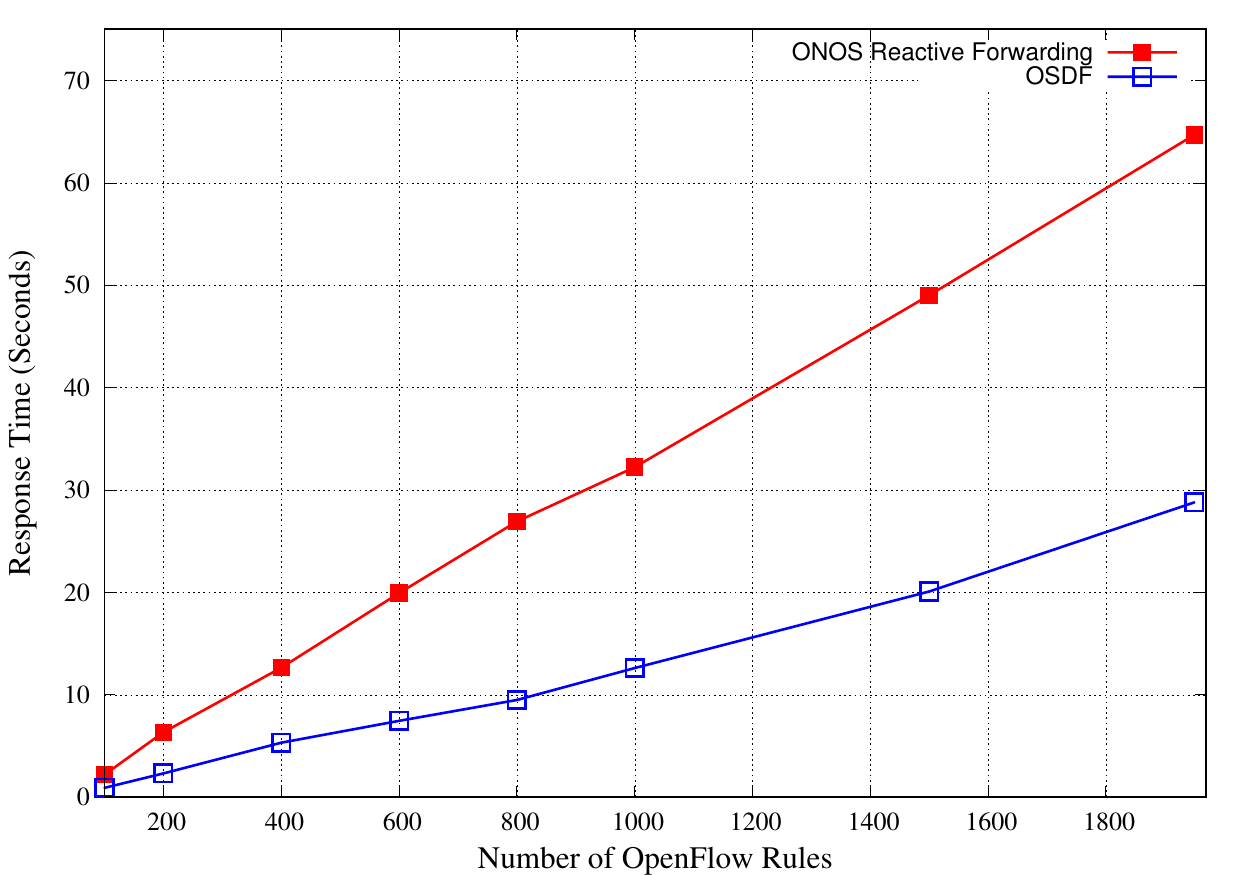}
\caption{Response time, ONOS reactive forwarding vs OSDF for the testbed}
\label{Fig16}
\end{figure}

\section{Discussion and Future Work}
\label{futurework}
Our project is in the first phases of the development and we plan to complete it by introducing new high level network operation services and new features. Some of our main goals are:

\begin{itemize}
\item Implement the same architecture in other SDN controllers such as Ryu \cite{Ref33} and OpenDayLight \cite{Ref34}.
\item Devise a new set of abstract operation services to support network service chaining and provide network function units such as \emph{Network Address Translators (NATs)}, load balancers, \emph{Intrusion Detection Systems (IDSes)}.
\item Design and implement a more scalable solution of the proposed architecture by distributing the policy storage among multiple instances of a SDN controller.    

\end{itemize}

\section{Conclusion}
This paper presents an SDN-based network programming framework which is called OSDF. OSDF provides a set of high level network service operations that can be invoked by management applications for network configuration and monitoring. OSDF uses a high-level approach  that allows a network administrator to enter high level network policies into the system without specifying any details about low level match-action fields. When a packet arrives, OSDF reactively examines policies, derives the required low-level flow rules, and installs flow entries in network switches along the path the packet will take. One of our key contributions is that we provide a higher level of abstractions than previous network programming languages.  Our system allows a manager to specify application requirements without giving layer 2 and 3 packet header fields. OSDF is equipped with a policy conflict management module that examines a set of policies, detects conflicts, and uses a conflict resolution algorithm to provide high level suggestions to the network administrator about ways to resolve the conflicts. We used OSDF to implement typical SDN applications such as an Inbound Traffic Engineering system, Inter-domain rate limiting, and Traffic Isolation. OSDF also can be used to implement other SDN applications such as firewall, application specific peering, intra-domain rate limiting.  In addition, we optimized flow rule installation by installing all of the required flow rules for a path when the first packet for the path arrives at a controller. Our simulation results show that the optimization reduces the negative impact of reactive part on the response time.

\label{conclusion}

\IEEEpeerreviewmaketitle

\bibliographystyle{IEEEtran}
\bibliography{ref2}

\end{document}